\documentstyle[12pt,aasms4]{article}

\slugcomment{Submitted to Astrophys. J.}

\lefthead{Frost, Lattanzio and Wood}
\righthead{Degenerate Thermal Pulses}

\newcommand{\msun}{M_{\sun}}

\begin{document}

\title{Degenerate Thermal Pulses in AGB Stars}

\author{C. A. Frost}
\affil{Department of Mathematics, Monash University, Clayton, 
Victoria~3168, Australia}

\author{J. C. Lattanzio}
\affil{Department of Mathematics, Monash University, Clayton, 
Victoria~3168, Australia} \and
\affil{Laboratoire d'Astrophysique, Observatoire de Grenoble, 
University Joseph Fourier, BP 53, F-38041, Grenoble Cedex 9, France}

\and

\author{P. R. Wood}
\affil{Mount Stromlo and Siding Spring Observatories, Private Bag, Weston
  Creek PO, ACT~2611, Australia}



\begin{abstract}
We report on the discovery of a new kind of thermal pulse in intermediate mass
AGB stars. Deep dredge-up during normal thermal pulses on the AGB leads to
the formation of a long, unburnt tail to the helium profile. Eventually the
tail ignites under partially degenerate conditions producing a strong
shell flash with very deep subsequent dredge-up.  The carbon content of the
intershell convective region (X$_{C}$ $\sim$ 0.6) is substantially higher than
in a normal thermal pulse (X$_{C}$ $\sim$ 0.25) and about 4 times more carbon is
dredged-up than in a normal pulse.
\end{abstract}




%

\section{Introduction}

It is now just over 30 years since thermal pulses were discovered in AGB stars
by \cite{Sch65} and over 20 years since 
third dredge-up was discovered by \cite{Ibe75}, 
and we are still learning about the
consequences of these events for stellar evolution and nucleosynthesis.
Although much is known  there are still many uncertainties,
especially concerning third dredge-up (\cite{Fro96}) and mass-loss.
We are presently studying the effects of hot bottom burning (HBB)
on intermediate mass AGB stars during their thermally pulsing
evolution. During these calculations we found a new kind of thermal 
pulse which we report on in this paper.

\section{Degenerate Thermal Pulses}

\subsection{Setting the Scene:~Deep Dredge-Up}

We consider the evolution of a $5\msun$
star with $Z=0.004$, as calculated 
with an updated version of the Mount~Stromlo Stellar Structure
Program. Opacities are taken from \cite{OPAL}, the mass-loss rate is from
\cite{VW93}\footnote{Except that we do not include the modification for
$M>2.5\msun$, which reduces the mass-loss rate for these stars.}, 
and we treat dredge-up in the way described by \cite{Fro96}.

\begin{figure}
\plotone{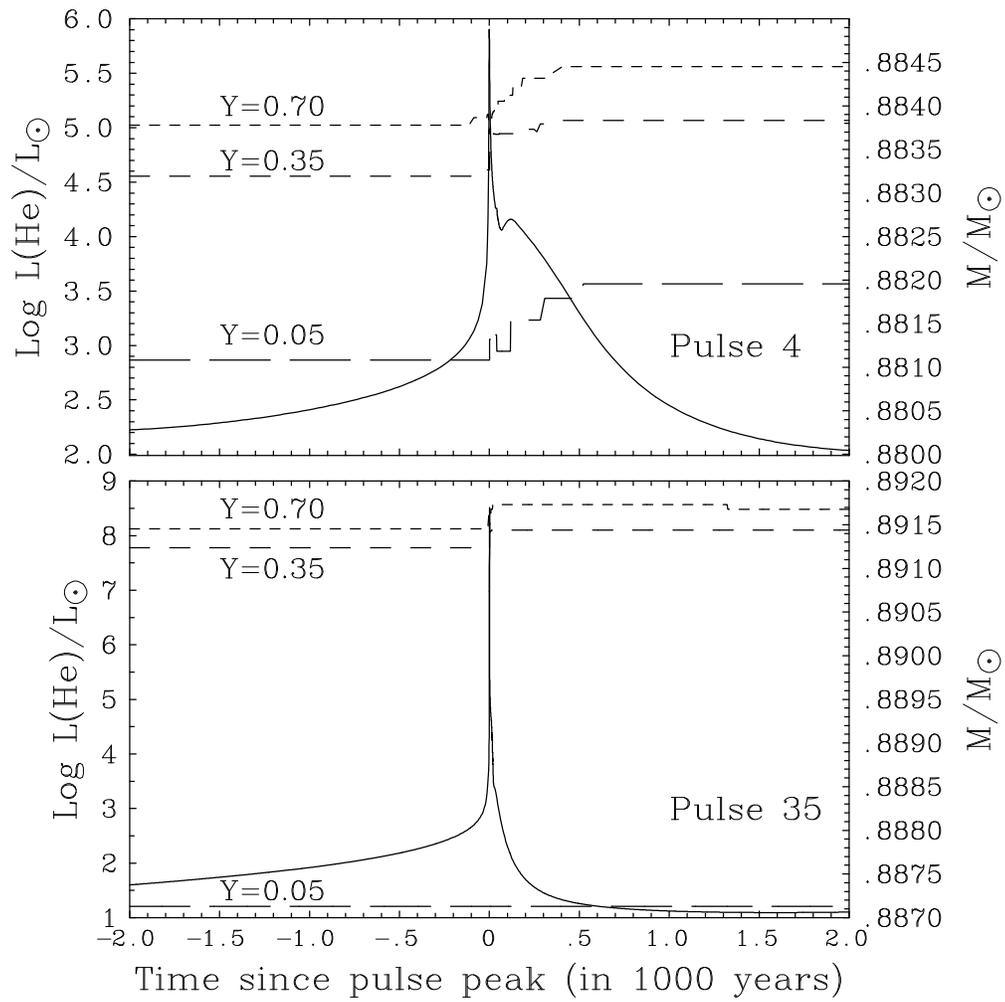}
\caption{The time variation of the helium luminosity and the position of the
  three points within the helium shell, at Y $=0.05$, $0.35$, and $0.70$. The upper
  panel shows the results for pulse 4 and the lower panel is for pulse 
  35.\label{Lsubpulse}}
\end{figure}

\begin{figure}
\plotone{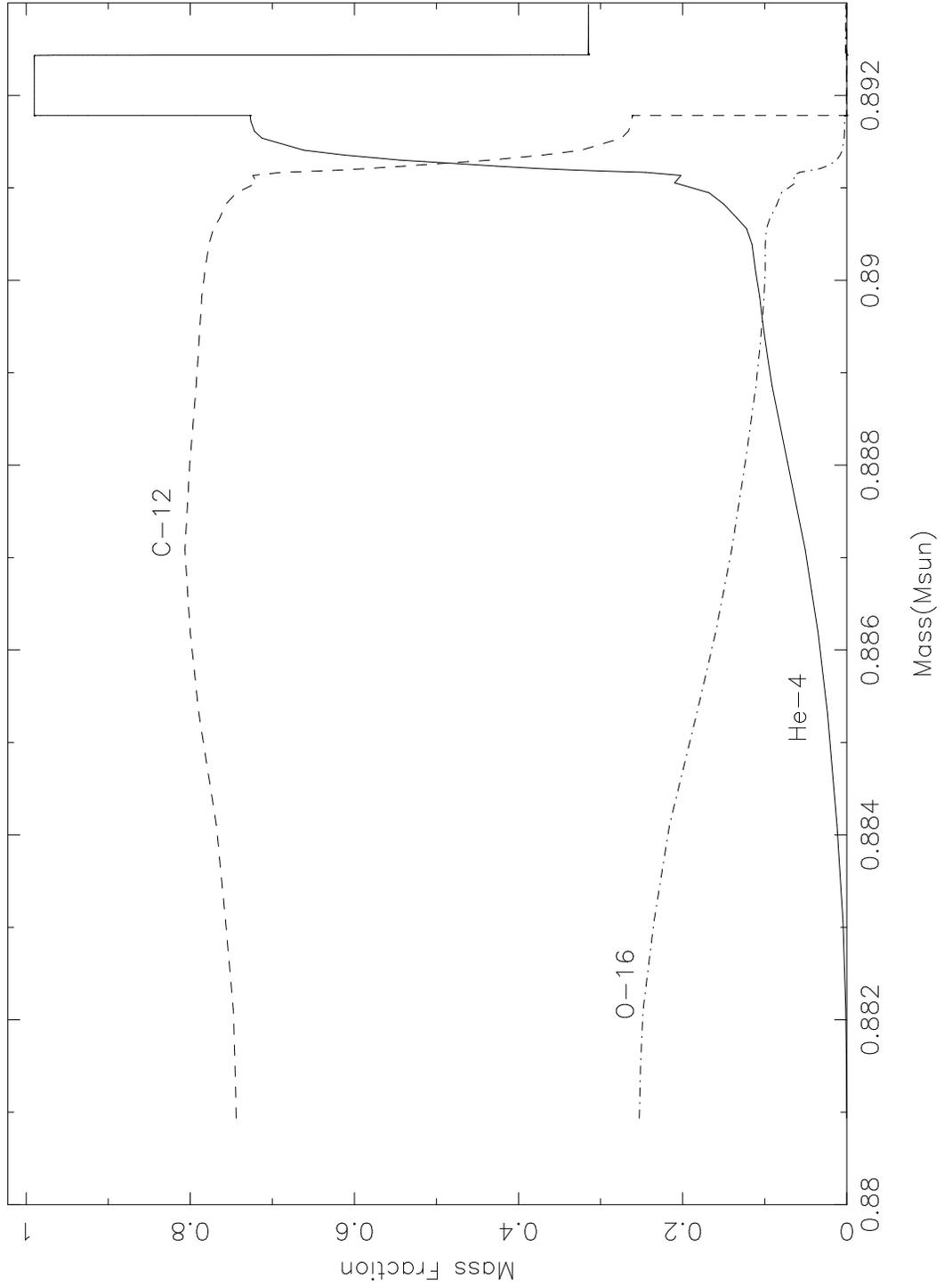}
\caption{Abundance profile of the helium shell after the 34$^{th}$ pulse.
  Note the long helium tail.\label{hetail}}
\end{figure}

\begin{figure}
\plotone{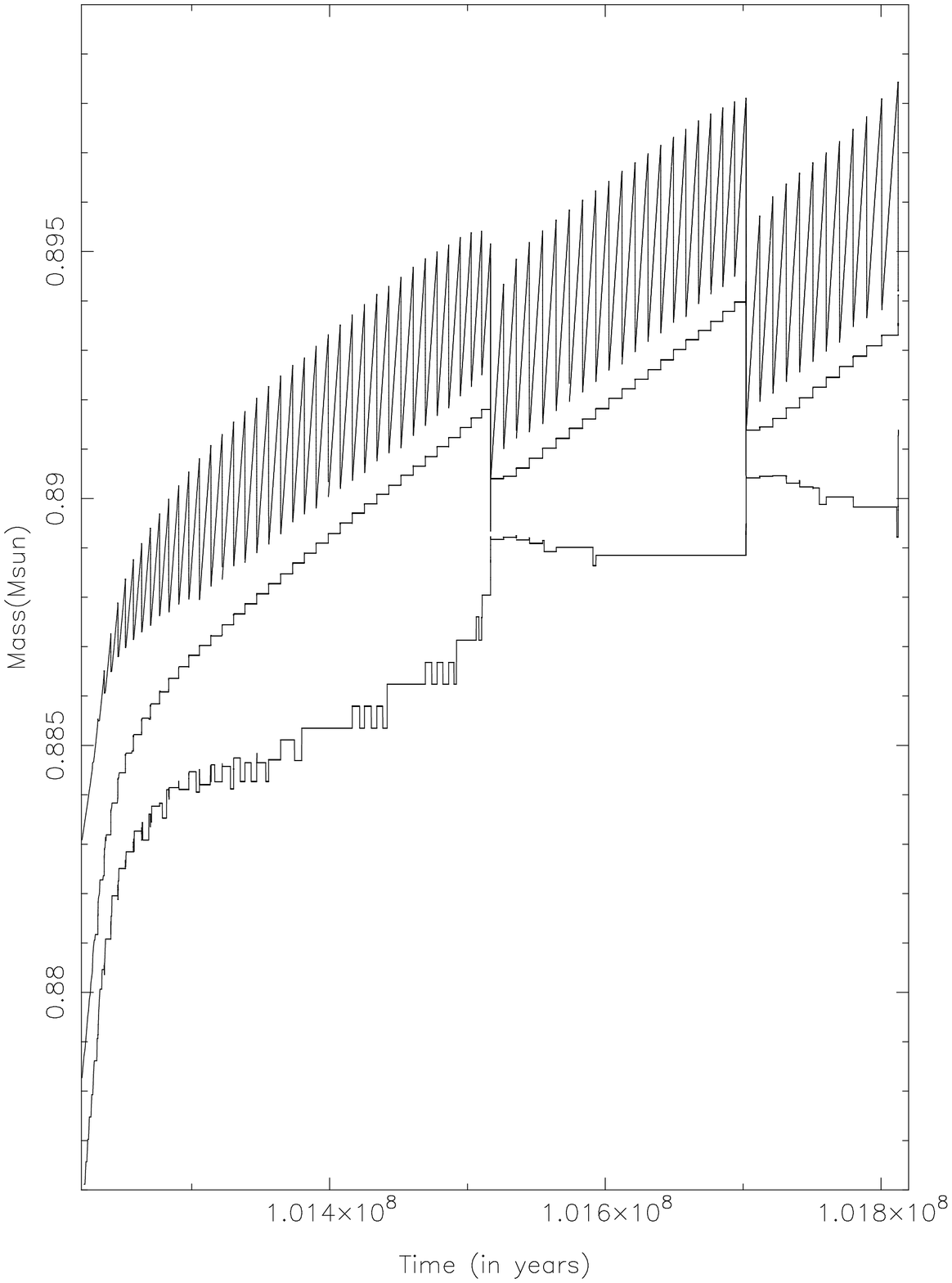}
\caption{The time variation of various important positions (in mass) within the
  model. From top to bottom these are: the hydrogen shell,
  the centre of the helium shell (where $Y=0.35$) and 
  the bottom of the helium shell (where $Y=0.05$). 
  \label{mass-time}}
\end{figure}

The star evolves as expected for the first 36 thermal pulses. We find
very deep dredge-up, as also found by \cite{VW93}, with the dredge-up
parameter $\lambda$ reaching close to unity after the first few pulses.
This turns out to be crucial:~immediately after a pulse, the deep
dredge-up cools the intershell region and does not allow the helium shell to
burn outward as much as it does when $\lambda$ is smaller. This is shown in
Figure~\ref{Lsubpulse} where we plot the time variation of the 
helium luminosity and the
position of three representative points within the helium shell. The upper
frame of the figure shows
the case for the 4th pulse, when $\lambda=0.78$. 
The period of extended helium burning following the pulse peak will be
referred to 
as the ``subpulse'', and the figure shows clearly that 
it is primarily during
the subpulse that the helium shell advances outward in mass. The lower frame
of the figure shows the same situation for the 35$^{th}$ pulse, when
$\lambda=0.93$. Here we see that the deep dredge-up extinguishes helium
shell burning very quickly, with the result that the shell moves outward
only a
small distance in mass. In particular, note that the bottom of 
the shell, at low helium values, hardly moves at all. 
This results in the helium shell developing a very long `tail' with
$Y \lesssim 0.1$, 
as shown in Figure~\ref{hetail}. 
Note that this tail develops only when the dredge-up becomes very deep, 
typically for $\lambda \gtrsim 0.8$.  In summary, 
during the first few pulses the advance in mass
of the helium shell matches
that of the hydrogen shell. However, once $\lambda$ approaches unity the
bottom of the helium shell starts to fall behind the rest of the shell,
thereby widening it in mass. These developments are clearly illustrated 
in Figure~\ref{mass-time}.

\begin{figure}
\plotone{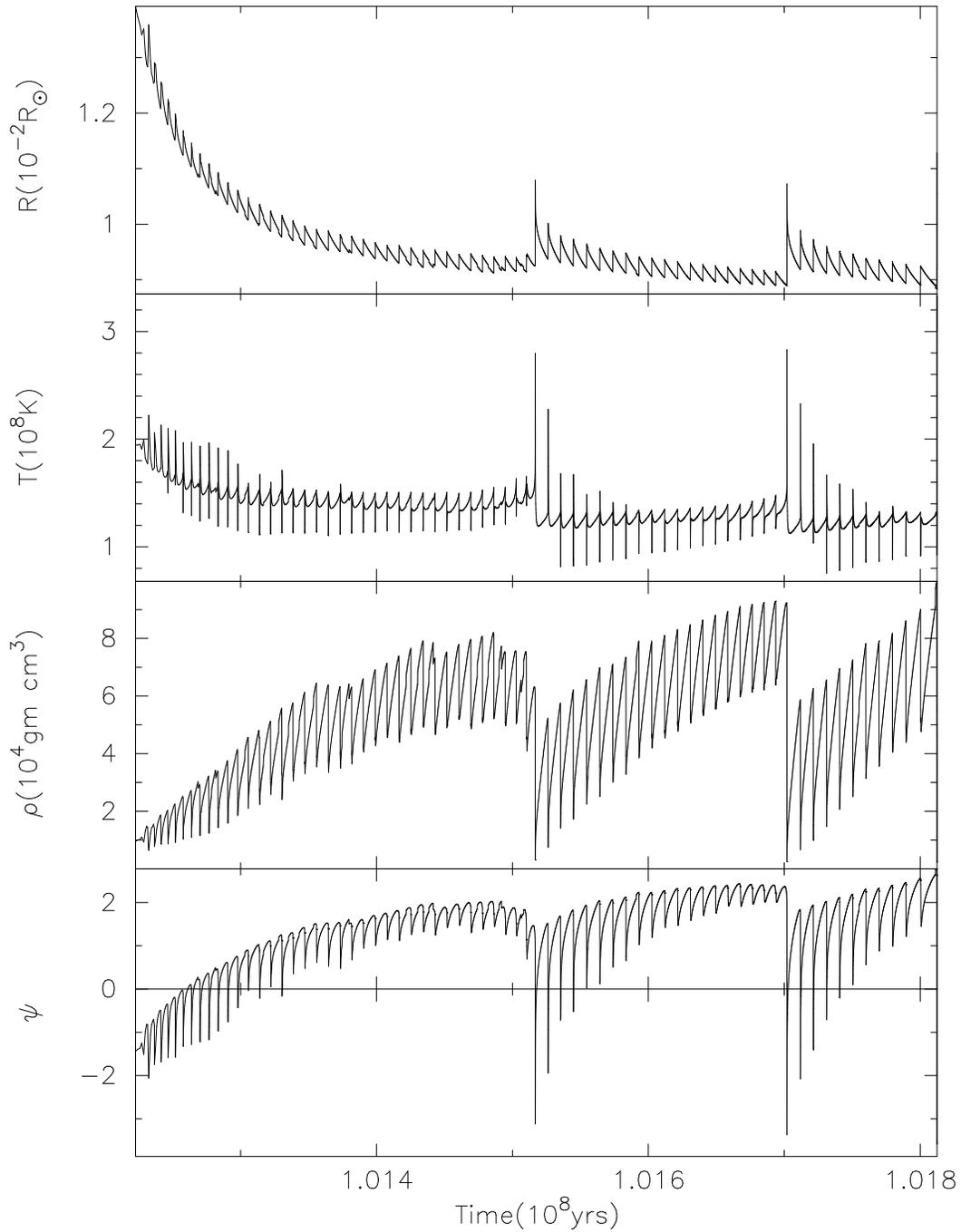}
\caption{Radius, temperature, density and degeneracy parameter at the base
  of the helium shell (where Y $=0.05$) as functions of time.
  Note that the earliest pulses (and the first few following a degenerate pulse)
  show a brief rise in temperature. This is extinguished once deep dredge-up
  develops.
\label{dpconds-time}}
\end{figure}

With successive pulses both the density and degeneracy at the shell base rise.
This is shown in Figure~\ref{dpconds-time}, which
presents conditions at the point where the helium abundance is 0.05.
Because of the cooling effect of dredge-up, the temperature
experienced by the long tail of the helium shell is relatively low and
helium does not burn (during the subpulse)
as it normally does in the absence of deep dredge-up.
Figure~\ref{dpconds-time} shows that each of the early pulses 
(with low $\lambda$)
experiences a brief time at high temperature, as represented by the temperature 
spikes. At this time the helium in
the shell is usually burned, and the helium shell advances. But later pulses
do not experience the brief high-temperature phase, and the helium tail
begins to develop.
Throughout all these phases, the density continues to rise as the core
contracts. After a few tens of pulses, the 
tail of the helium shell extends down to densities
not usually encountered
by the helium shell in AGB stars and the bottom of the helium shell is  
mildly degenerate.

\subsection{A Degenerate Pulse}

\begin{figure}
\plotone{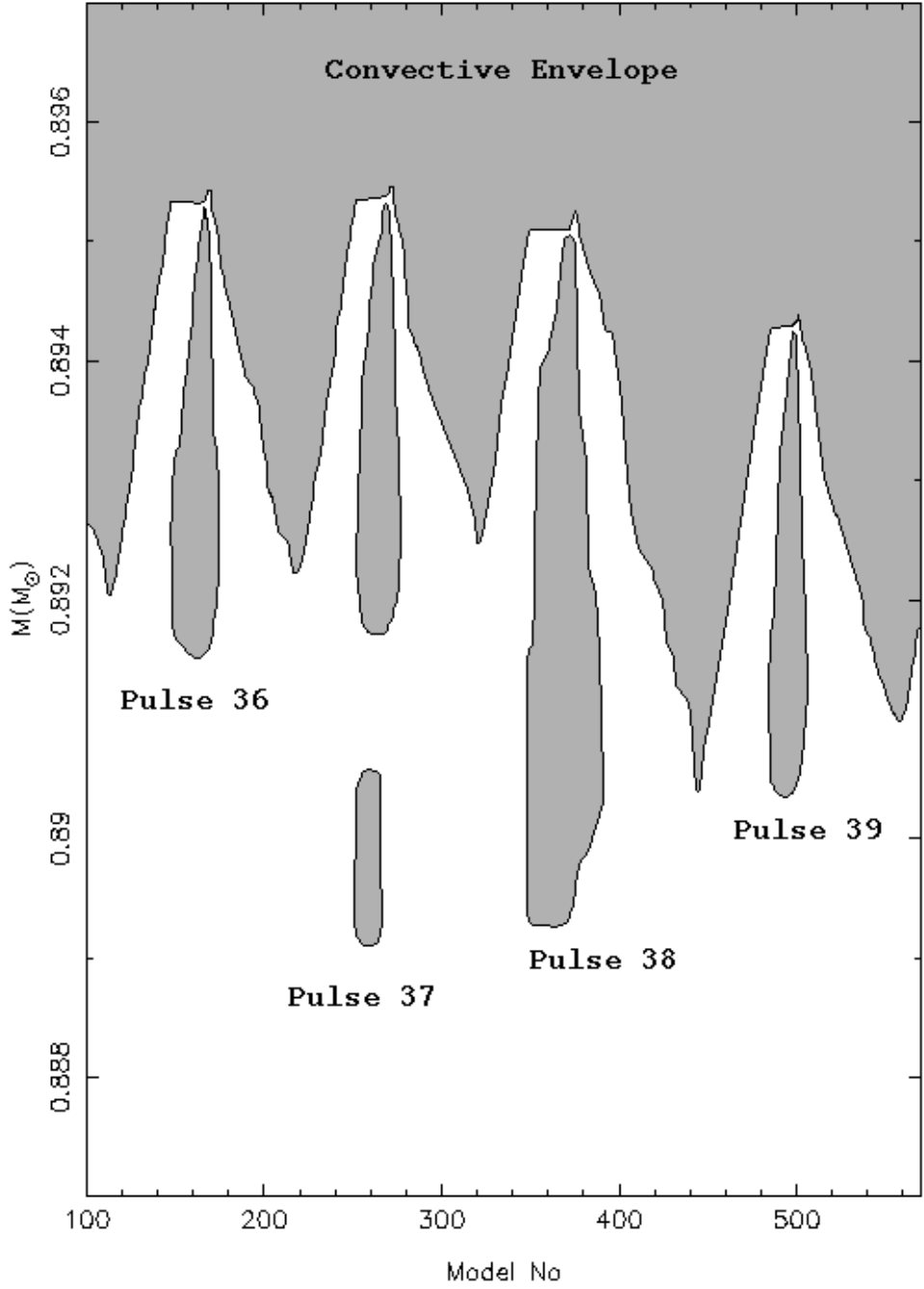}
\caption{Convective zones versus model number for pulses 36--39 of the model
  described in the text. Note that the $x$-axis is actually the model number
  in a post-processing nucleosynthesis code, {\bf not} the evolution model
  number.\label{dpconv-time}}
\end{figure}

During the 38$^{th}$ thermal pulse, the tail of the helium shell ignites.
The peak of the burning is now centered at the
base of the shell, rather than near the shell center, which is the normal
case. This results in a very deep convective shell which extends from the 
{\it bottom\/} of the helium shell almost to the hydrogen shell, as
clearly illustrated in
Figure~\ref{dpconv-time}, which shows the 
convective boundaries during pulses 36--39. The isolated
convection zone seen in Figure~\ref{dpconv-time} 
at the bottom of the normal intershell convection zone in
pulse 37 will be discussed below.

Thermal pulse number 38 occurs at higher temperatures and is greater in
strength than normal.  The abundances of 
$^4$He and $^{12}$C left by the intershell
convective zone are about 0.34 and 0.58, respectively, compared with 
0.74 and 0.24 in the previous pulse.  Thus a large quantity of $^{12}$C
has been produced. Large expansion is driven by the pulse and very deep 
dredge-up occurs.  The extent
of dredge-up ($\lambda \approx 1.8$) is such that roughly 80\% of the carbon
is dredged-up by the envelope.  {\it Nearly four times the normal mass of carbon is
dredged-up in a single pulse!}  The surface ratio of carbon to oxygen rises
from 0.36 at the beginning of the pulse to 1.04 at the end, creating a 
carbon star in a model that will undergo significant Hot Bottom
Burning during the 
next interpulse
phase.

\subsection{A Failed Degenerate Pulse}

Another feature seen in Figure~\ref{dpconv-time} is a small convective
region at the
bottom of the usual
intershell convective zone for the 37$^{th}$ pulse
(at $M_r\sim 0.890\msun$). This is exactly the same
phenomenon we have discussed above, but in this case the flash was not
strong enough for the convective zone at the base of the shell to grow and
engulf the usual intershell convective zone. This event
is really a failed degenerate pulse. The amount
of helium burned is quite small.

\subsection{Degenerate Sub-Pulses}
\begin{figure}
\plotone{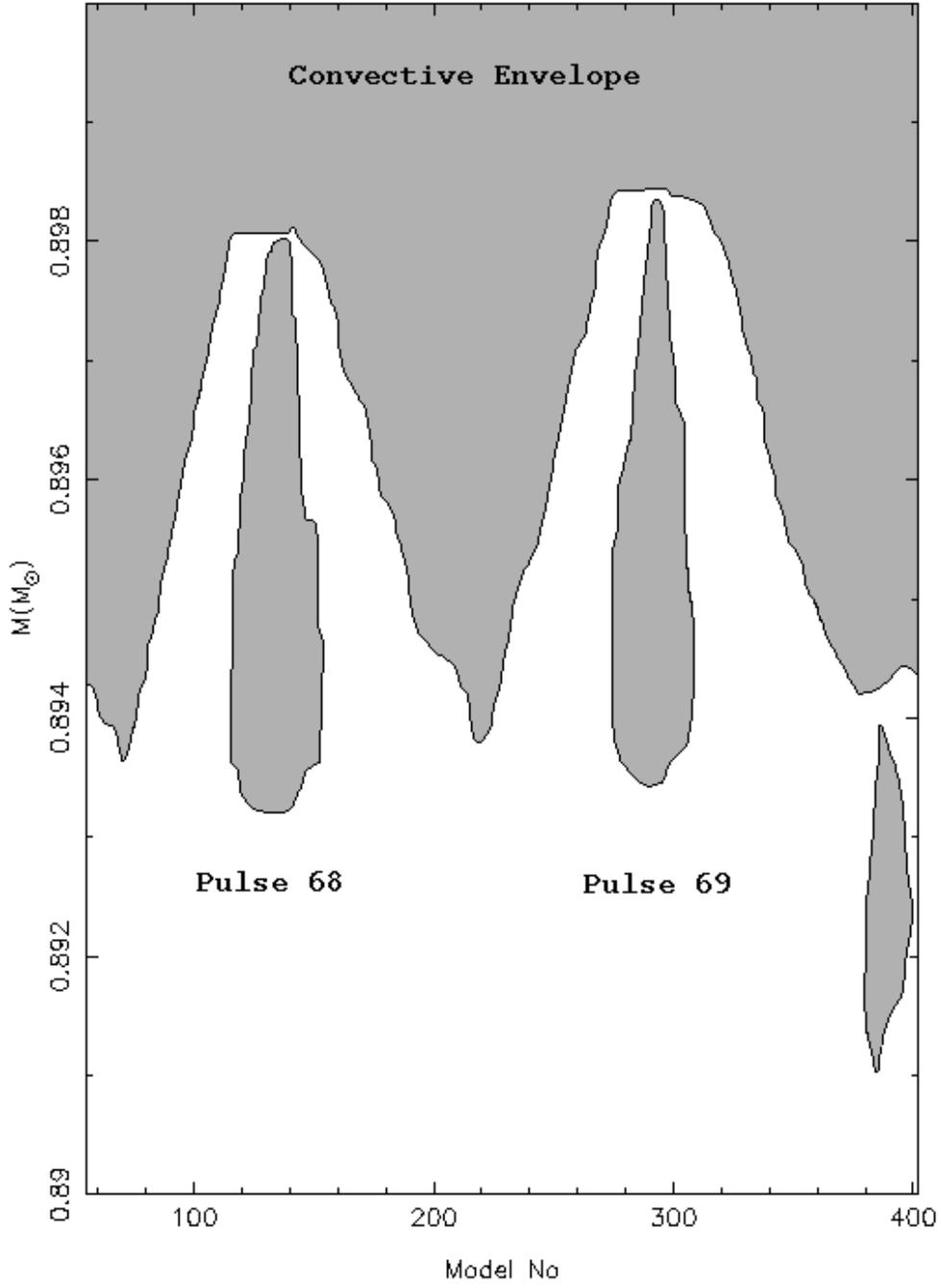}
\caption{Convective zones versus model number for pulses 68 and~69.
  \label{dpsubconv-time}}
\end{figure}

After the degenerate 38$^{th}$ pulse, the helium shell is reduced in
width and is now closer to a normal position found when dredge-up is
shallower. Subsequent evolution, shown in Figures~\ref{mass-time} 
and~\ref{dpconds-time}, shows that
as deep dredge-up continues, the same situation develops once more. The
helium shell widens because the low helium tail of the shell remains 
unburned after
each pulse, and a second degenerate pulse occurs 20 pulses
later at pulse~58. Following
this we again find the helium shell widening, but the next degenerate pulse
occurs after only 11 more pulses.  This degenerate pulse is different:
it occurs {\it during the dredge-up 
phase} of the 69$^{th}$ pulse as shown in
Figure~\ref{dpsubconv-time}. 
The development of this ``subpulse'' is due to a complex
interplay between helium burning during the subpulse and the cooling
produced by envelope convection, resulting in a small helium
inversion in the tail of the helium shell.

\subsection{Other Occurrences of Degenerate Pulses}
During the calculation of nine evolutionary sequences ($M=4$, $5$ and
$6\msun$, each
with $Z=0.004$, $0.008$ and $0.02$) we  found six
degenerate pulse events. Three occurred during the evolution of the $5\msun$
$Z=0.004$ case described here.
One occurred for the $5\msun Z=0.008$ case, and
two during the evolution of the $6\msun$ model with $Z=0.004$. It would
certainly be premature of us to claim to have isolated any critical ranges
in mass or composition for these events.

\section{Physical and Numerical Dependence}

We propose that degenerate pulses and subpulses are caused by extremely
deep dredge-up cooling the helium shell, thereby preventing burning of 
the lower helium
shell which normally occurs during the subpulse phase.  A long tail (in mass)
of low helium abundance ($Y\lesssim 0.1$) forms and grows increasingly denser
until finally it ignites under mildly
degenerate conditions, either at the beginning
of a pulse (a degenerate pulse) or during dredge-up (a degenerate 
subpulse).

\subsection{The effect of the entropy of mixing}

\begin{figure}
\plotone{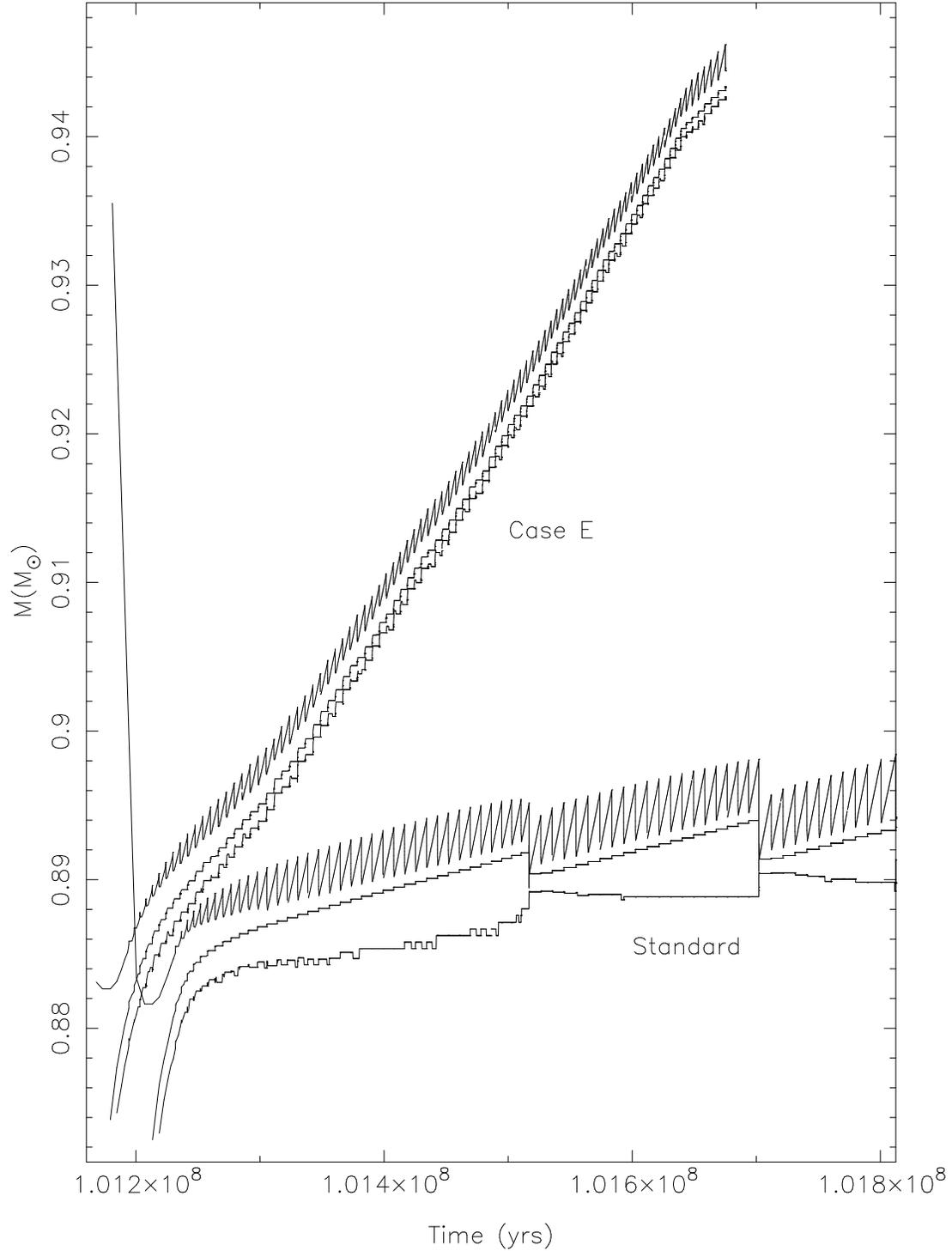}
\caption{Core mass, position of the middle of the helium shell ($Y=0.35$),
  and bottom of the helium shell ($Y=0.05$) for both the standard case
  (with high $\lambda$) and Case~E (with lower $\lambda$).
  \label{case-E}}
\end{figure}

To test the effects which  the numerical treatment of dredge-up and the 
entropy of mixing have on the occurrence of degenerate pulses
we evolved
a 5$\msun$ $Z=0.004$ star from the main-sequence through the thermally
pulsing AGB evolution
with Wood's (1981) entropy treatment removed (hereafter Case E).
This reduced $\lambda$ from an average of 0.95 to
0.67.  
Figure~\ref{case-E} shows the time variation of the
mass of the H-exhausted core
together with masses at the middle and bottom of the helium shell.
Also plotted on the same figure for comparison are the shell positions for
our standard $M=5\msun$ $Z=0.004$ model (discussed above) 
with deep dredge-up and degenerate pulses.
No degenerate pulses occur in Case~E.  
We see in this Figure
that the tail of the helium shell burns during each pulse
and the shell remains relatively thin in mass.  Thus pulses
proceed normally.  

To test the possibility that the entropy treatment itself may affect the
occurrence of degenerate pulses, we evolved another $5\msun$ $Z=0.004$
model with the treatment included but with $\lambda$ restricted to a maximum
value of 0.67.  $\lambda$ tended to
vary from pulse to pulse for this model, though never exceeding 0.67, and
the mean value was 0.54.  Otherwise the results were very similar to 
Case~E with the helium
shell remaining very narrow in mass and no degenerate pulses or subpulses
occurring.  

\subsection{The effect of time steps}

To test whether time-stepping influences the development of degenerate thermal
pulses we ran a case with twice the time resolution of our standard $5\msun$\
$Z=0.004$ model, and evolved it through 45 thermal pulses.  We saw all the
features described for the standard model 
and a degenerate pulse occurred during the $42^{nd}$ pulse.
Overall, this model is similar to the standard model, though the first
degenerate pulse occurs some 4 pulses later.  Clearly, the fine details of
stellar evolutionary calculations are dependent on numerical treatment,
including time-stepping (see also \cite{str97}).  The important result of this
test sequence is that reducing the time steps makes only a small difference to
the behaviour of the helium shell when deep third dredge-up is present.

\subsection{The effect of spatial zoning}

In Figure~\ref{mass-time} we showed the variation with time of the position (in
mass) of the hydrogen and helium shells of the standard $M=5\msun$ $Z=0.004$
model.  The reported positions of the helium shell base in
Figure~\ref{mass-time} are somewhat ragged.  This is due to the method used to
define the mass interior to $Y$ = 0.05 in the model: we select the point ``at''
$Y=0.05$ by selecting the shell $j$ so that $Y(j) < 0.05$ and $Y(j+1) > 0.05$.
In addition to the raggedness of the position of the point at $Y$ = 0.05, the
reported position of the shell bottom apparently moves slowly inward following
the two degenerate pulses.  This is a non-physical effect due to numerical
diffusion as model zoning is changed.  Both the above features indicate that
the mesh spacing may not be fine enough at the bottom of the helium shell.

To test whether mesh spacing
influences the occurrence of degenerate pulses, we evolved another
$5\msun$\ $Z=0.004$ star with exactly the same 
conditions as the standard case,
except that mesh spacing was halved everywhere.
This star experienced a total of 65 complete pulses before convergence
difficulties terminated calculations during the $66^{th}$ pulse.  A degenerate
pulse occurred during the $40^{th}$ pulse, and a degenerate subpulse during
the $51^{st}$ pulse.

\begin{figure}
\plotone{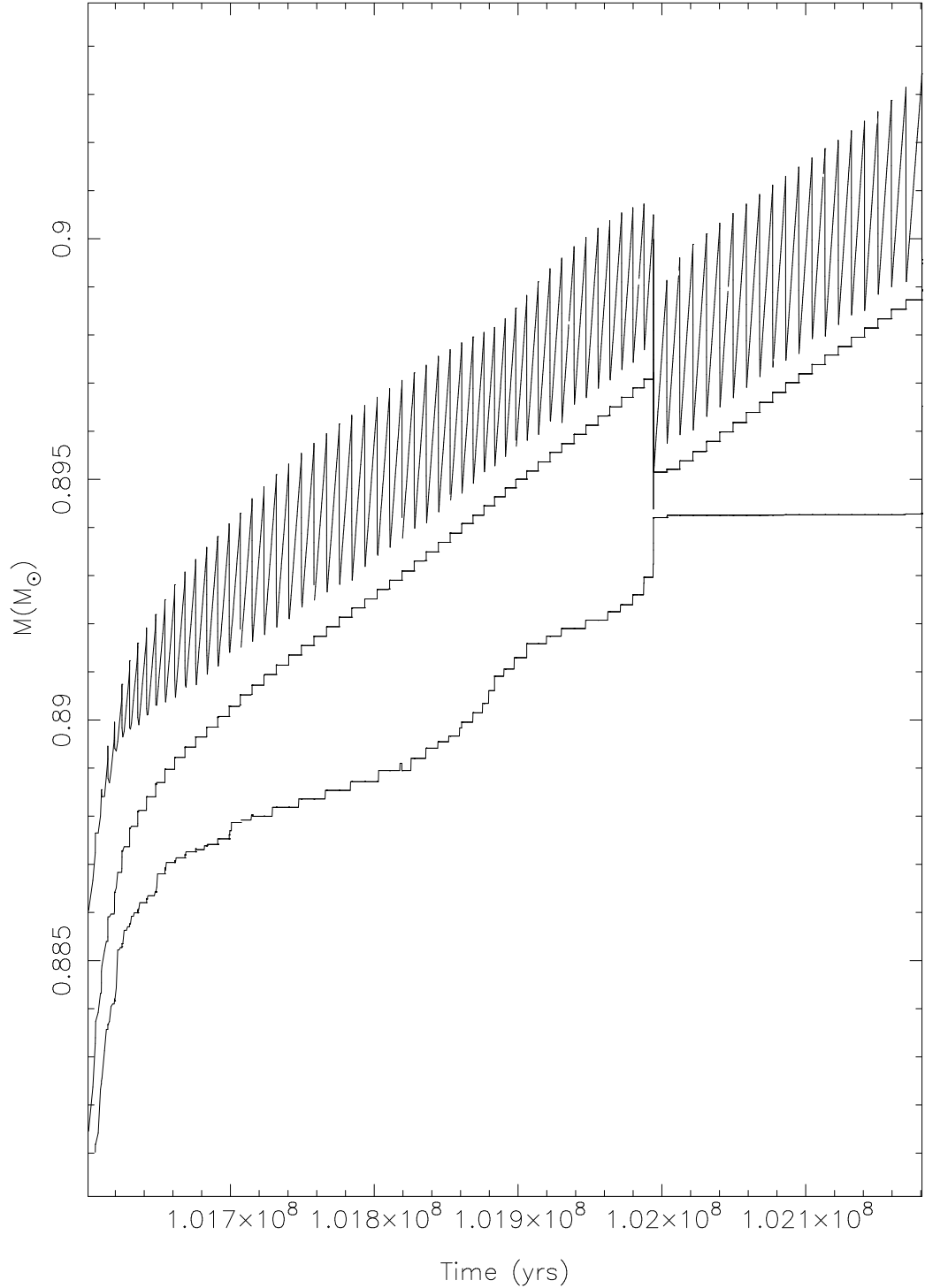}
\caption{The time variation of various important positions (in mass) within the
  model. From top to bottom these are: the hydrogen shell,
  the centre of the helium shell (where $Y=0.35$) and 
  the bottom of the helium shell (where $Y=0.05$). This should be compared with
  the lower resolution calculation in Figure~\ref{mass-time}.
  \label{figmeshshells.ps}}
\end{figure}

\begin{figure}
\plotone{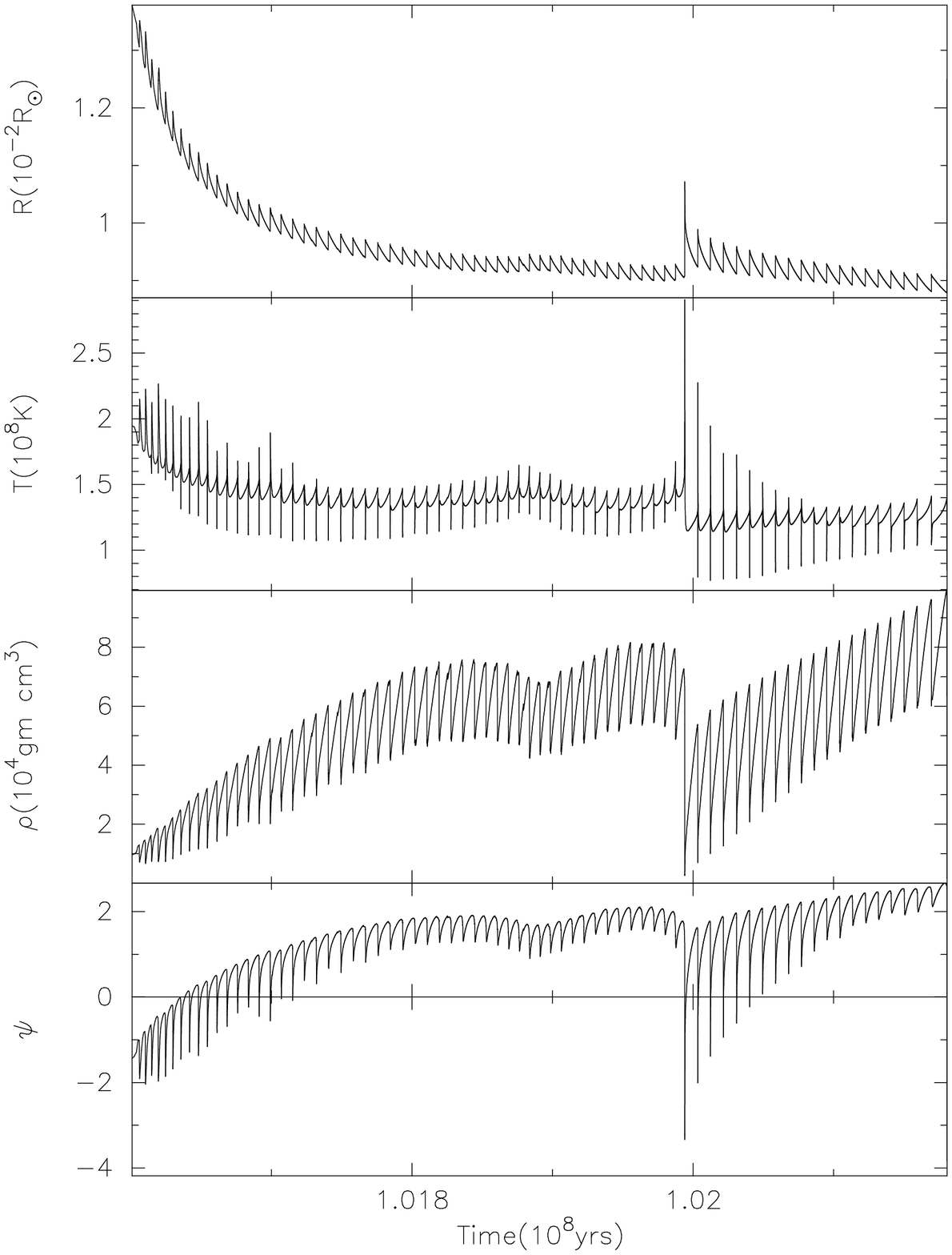}
\caption{Radius, temperature, density and degeneracy parameter at the base
  of the helium shell (where Y $=0.05$) as functions of time.
  This should be compared with
  the lower resolution calculation in Figure~\ref{dpconds-time}.
  \label{figmeshm3.ps}}
\end{figure}

As a further test, we
evolved another case with mesh spacing returned to normal
throughout the entire model except at the base of the helium shell. 
Between $Y=0$ and 0.15, the mesh spacing was made 10 times finer than normal.
The results are shown in Figures~\ref{figmeshshells.ps} and \ref{figmeshm3.ps}.
The usual thickening of the helium
shell occurred,
but once the pulse number exceeded 30, the base
of the helium shell began to move outwards again, the
temperature rose at the base, and the density and degeneracy
fell.  The star barely avoided a degenerate pulse at this time, with the base of the 
helium shell burning gently.
However, the easing of conditions at the shell base was short-lived and a 
degenerate pulse finally occurred at pulse 52.  Afterwards,
the base of the helium shell remained essentially
stationary in mass.  Consequently, the helium shell thickened again 
and a degenerate subpulse was
triggered at the end of the $72^{nd}$ pulse, where the
calculations were terminated. 

We conclude from the above tests that the details, {\it but not the 
occurrence,\/} of
degenerate pulses may be influenced by numerical details.  Degenerate
pulses inevitably
occur if dredge-up is deep enough for long enough.

\section{Conclusion}
We believe that the development of degenerate thermal pulses is
inevitable provided two criteria are met:~1)~dredge-up is very deep,
providing cooling of the helium shell soon after ignition;~and~2)~the 
star lives
long enough on the AGB for the helium shell to reach 
(partially) degenerate conditions. 
Whether these
conditions are realised in real stars remains an open question. The two
aspects of AGB evolution which are the most uncertain are the 
extent of dredge-up and
the rate of mass-loss (hence AGB lifetime), 
precisely the two phenomena which govern the occurrence
or not of degenerate thermal pulses.

It seems prudent to look for some observational consequence of 
degenerate thermal pulses. The nucleosynthetic consequences are currently
being investigated, and will be reported elsewhere.

\clearpage

\clearpage


%

\end{document}